\documentclass[11pt,tightenlines,eqsecnum,floats,aps,amsmath,amssymb,nofootinbib,prd,shownopacs,floatfix]{revtex4}

\usepackage{graphicx}
\usepackage{epstopdf}
\usepackage{latexsym}
\usepackage{amssymb}
\usepackage{amsmath}
\usepackage{color}
\usepackage{mathrsfs}
\usepackage{xparse}
\usepackage{float}
\usepackage{mathtools}
\usepackage{multirow}
\usepackage[center]{subfigure}

\usepackage{xcolor}
\begin{document}
\renewcommand\arraystretch{2}
 \newcommand{\bq}{\begin{equation}}
 \newcommand{\eq}{\end{equation}}
 \newcommand{\bqn}{\begin{eqnarray}}
 \newcommand{\eqn}{\end{eqnarray}}
 \newcommand{\nb}{\nonumber}
 \newcommand{\cb}{\color{blue}}
    \newcommand{\cc}{\color{cyan}}
     \newcommand{\lb}{\label}
        \newcommand{\cm}{\color{magenta}}
\newcommand{\rc}{\rho^{\scriptscriptstyle{\mathrm{I}}}_c}
\newcommand{\rd}{\rho^{\scriptscriptstyle{\mathrm{II}}}_c}
\NewDocumentCommand{\evalat}{sO{\big}mm}{%
  \IfBooleanTF{#1}
   {\mleft. #3 \mright|_{#4}}
   {#3#2|_{#4}}%
}
\newcommand{\PRL}{Phys. Rev. Lett.}
\newcommand{\PL}{Phys. Lett.}
\newcommand{\PR}{Phys. Rev.}
\newcommand{\CQG}{Class. Quantum Grav.}
\newcommand{\parallelsum}{\mathbin{\!/\mkern-5mu/\!}}

\title{Loop Quantum Gravitational Signatures via Love Numbers}

\author{Meysam Motaharfar$^1$}
\email{mmotah4@lsu.edu}
 \author{Parampreet Singh$^{1,2}$}
\email{psingh@lsu.edu}
\affiliation{$^1$ Department of Physics and Astronomy, 
Louisiana State University, Baton Rouge, LA 70803, USA}
 \affiliation{$^2$ Center for Computation and Technology,
Louisiana State University, Baton Rouge, LA 70803, USA}

\begin{abstract}
Loop quantum gravitational effects can resolve the central singularity of black holes while potentially leaving tiny traces of quantization in the exterior spacetime. We show the way these residues can, in principle, be explored using tidal Love numbers (TLNs). We consider loop quantized  Schwarzschild black hole, in particular the Ashtekar-Olmedo-Singh (AOS) model, and study the static response to external tidal fields of spin zero (scalar field), spin one (vector field), and spin two (axial gravitational field) types. We find that, in contrast to the classical theory, where TLNs vanish, they are non-vanishing and negative for all three responses and for all multipoles. Besides, the magnitude of  TLNs decreases as the black hole mass increases, and  TLNs, in response to the axial gravitational field, have the largest magnitude among these three responses. Our results show that for black holes of mass $M \gtrsim 4.3 \times 10^{4} M_{\textrm{Pl}}$, the AOS model is consistent with current and next-generation detection limits for TLNs. Our findings suggest that the quantum deformability of loop quantum black holes, arising from the inherent fuzziness of spacetime geometry, reveals a fundamentally distinct internal structure compared to their classical counterparts. This unique feature manifests as quantum hair, which, in principle, can be detected by future observations.
\end{abstract}

\maketitle

\section{Introduction} 

Black holes in general relativity (GR) are remarkably simple, uniquely defined by only a few external macroscopic parameters: mass, angular momentum, and electric charge. This inherent simplicity also extends to the linear and static response of black holes to external tidal fields, strengthening the so-called no-hair theorem \cite{no-hair} beyond the background geometry.
With vanishing static tidal Love numbers (TLNs) \cite{10.1093/mnras/69.6.476}, four-dimensional black holes in vacuum GR turn out to be not only the simplest but also the most rigid objects in nature. It has been demonstrated that TLNs for several families of asymptotically flat black holes in four-dimensional spacetime vanish. This includes, namely, Schwarzschild \cite{Schwarzschild, Kol:2011vg, Hui:2020xxx}, Kerr \cite{LeTiec:2020spy, Kerr}, Reissner-N\"ordstrom \cite{Cardoso:2017cfl, Pereniguez:2021xcj, Rai:2024lho}, and Kerr-Newman \cite{Ma:2024few} black holes (see Refs. \cite{non-linear TLNs, Combaluzier-Szteinsznaider:2024sgb, Gounis:2024hcm} for extensions to non-linear TLNs and also Refs. \cite{dynamical TLNs} for recent studies on dynamical TLNs). However, this peculiar property of vanishing TLNs is specific to four-dimensional GR, indicating the existence of hidden (or ladder) symmetries for spherically symmetric spacetimes in four-dimensions \cite{symmetry of TLNs, Combaluzier-Szteinsznaider:2024sgb, Gounis:2024hcm}. On the other hand, TLNs are non-vanishing for black holes in higher-dimensional spacetimes \cite{Hui:2020xxx, Kol:2011vg, Ma:2024few, Pereniguez:2021xcj, higher dimension TLNs}, in the presence of a cosmological constant \cite{Nair:2024mya, Franzin:2024cah}, extended gravitational theories \cite{Cardoso:2017cfl, Cardoso:2018ptl, DeLuca:2022tkm, Barura:2024uog, Katagiri:2024fpn}, or with matter in the environment surrounding the objects \cite{Cardoso:2019upw, DeLuca:2021ite, DeLuca:2022xlz, Katagiri:2023yzm}. Interestingly, tidal effects are imprinted in gravitational waveforms at 5th post-Newtonian order for a compact binary inspiral \cite{Flanagan:2007ix, Vines:2010ca, Narikawa:2023deu}. While events detected by LIGO and Virgo are consistent with vanishing TLNs \cite{Narikawa:2021pak, Chia:2023tle}, future detection of non-vanishing TLNs can be a smoking gun for new physics beyond GR at the horizon scale or evidence for the presence of non-trivial astrophysical environments. Remarkably, LIGO, LISA, and the Einstein Telescope are expected to constrain TLNs with significantly greater precision, enhancing our ability to probe deviations from GR in the near future \cite{Puecher:2023twf}.

Despite its tremendous success, GR suffers from the existence of singularities \cite{Geroch:1968ut, Hawking:1970zqf}, which signal the limit of validity of Einstein's theory. It has long been expected that quantum gravitational effects might smooth out or resolve spacetime singularities, revealing new physics at the Planck scale. While the final picture from quantum gravity is still under development, in recent years, techniques from loop quantum gravity (LQG) have been successfully applied to understand the quantization of various cosmological \cite{Ashtekar:2011ni} and black hole \cite{Ashtekar:2023cod} spacetimes. The primary result of the loop quantization of all the spacetimes studied so far is that quantum geometric effects resolve the classical singularities, such as replacing big bang with a big bounce \cite{Ashtekar:2006rx,Ashtekar:2006wn}. While the quantum evolution is determined by a non-singular quantum difference equation, extensive numerical simulations reveal that it can be accurately captured by an effective spacetime description \cite{Diener:2014mia,Diener:2017lde}. Using this effective description, all the strong curvature singularities are generically resolved in isotropic and anisotropic spacetimes \cite{singularity-resolution}, including the Kantowski-Sachs spacetime \cite{singularity-ks}, which in the vacuum case is isometric to the interior of the Schwarzschild black hole. Since loop quantum effects for the Kantowski-Sachs cosmology can be rigorously explored using techniques of loop quantum cosmology (LQC), various models have been proposed for loop-quantized Schwarzchild black holes \cite{Ashtekar:2005qt, Modesto:2005zm, Chiou:2008nm, Dadhich:2015ora, Corichi:2015xia, Olmedo:2017lvt, Ashtekar:2018lag, Ashtekar:2018cay, Bodendorfer:2019nvy, Zhang:2020qxw, ElizagaNavascues:2022rof}, which have motivated other explorations in loop quantum black holes (see for e.g. \cite{LQBH}) and related approaches (see \cite{Rovelli:2024sjl} for an overview). 

To explore potential quantum geometric effects in TLNs, the first question is which model shall be considered from available quantizations. Note that due to quantization ambiguities, arising, for example, from different treatments of the Hamiltonian constraint, different loop quantizations of the same classical spacetime are in principle possible, all of which seemingly resolve central singularity. However, it turns out that only a few models satisfy certain consistency requirements, which include, for example, independence of quantum gravity effects from fiducial structures and recovery of GR at small spacetime curvatures \cite{Corichi:2008zb}.  In particular, out of several loop quantizations of the interior of the Schwarzschild black hole,  Ashtekar-Olmedo-Singh (AOS) model \cite{Ashtekar:2018lag, Ashtekar:2018cay} is favored as it satisfies all the consistency requirements,  and is free of all the limitations of the previous models \cite{Ashtekar:2023cod}. 
In the AOS model, the classical singularity at the interior of a black hole is replaced with a transition surface, a non-singular bounce, joining a trapped region to its past and an anti-trapped region to its future, extending the Schwarzschild interior to encompass what is interpreted as a white hole horizon. In particular, it has been found that for the AOS model, if one considers macroscopic black holes, all curvature scalar invariants have uniform upper bounds. In addition, quantum effects are negligible in the low curvature limit, and the results are independent of the infrared cut-off. In the AOS model, quantization has also been performed for the exterior, whose properties have been studied \cite{Ashtekar:2020ckv}.

The phenomenology of the AOS model has been investigated in Ref. \cite{Ashtekar:2020ckv}, including studies on its quasi-normal modes \cite{Daghigh:2020fmw, Gingrich:2023fxu} and shadow \cite{Devi:2021ctm}. Moreover, a large family of solutions with properties similar to the AOS model, where polymerization parameters are constant on dynamical trajectories, was found in Ref. \cite{Ongole:2023pbs}, and it was also extended to include the cosmological constant \cite{Feng:2024sdo}. However, the static response of loop-quantized black holes, specifically the AOS model, to an external tidal field and the extraction of TLNs has not been studied in the literature. Although loop-quantized black holes are in four dimensions, the quantum gravitational effects lead to deviation from GR at the Planck scale, replacing the interior of black holes with a non-singular core, and it is expected that TLNs will be non-vanishing \cite{Brustein:2020tpg, Kim:2020dif}, encoding information about the internal structure and the gravitational dynamics of loop-quantized black holes. This raises several intriguing questions: Do loop-quantized black holes deform under the influence of an external tidal field, and if so, what is the extent of their deformability? Could this deformation leave detectable imprints on future gravitational wave observations? Furthermore, is it possible to distinguish between different loop-quantized black holes based on their TLNs? Finally, can TLNs provide a means to constrain the quantum parameters of loop-quantized black holes?

In this analysis, we assume the validity of the AOS metric as the background and study the static response of the AOS model to an external tidal spin-$s$ field, namely a scalar field, a vector field, and an axial gravitational field that act as perturbations. We extract TLNs in a closed form by expanding the exact solution of the radial component of perturbations at large distances. We find that TLNs depend on the spin of the external tidal field $s$, the multipole $\ell$, and also on a quantization parameter, one of the polymerization parameters of the model, namely $\delta_{b}$, which is proportional to $M^{-1/3}$ where $M$ is the black hole mass. Our results reveal that all scalar, vector, and tensor TLNs are non-vanishing and negative for all multipoles due to quantum gravitational effects, even in four dimensions (except the scalar TLN for $\ell=0$, which vanishes). We also observe that TLNs for the AOS model have a larger magnitude for larger quantum parameters, meaning that the black holes with smaller mass have larger TLNs. Moreover, the larger the multipole, the smaller the magnitude of TLNs for the same black hole mass. We also find that the tensor TLNs have the largest magnitude among these three responses. Considering the potential lower bound for detectable quadrupole TLN, which is roughly around $\kappa_{\ell=2} \simeq 2\times 10^{-5}$ \cite{Pani:2019cyc}, and requiring that the AOS model predicts negligible TLNs to be consistent with GR, we find that the AOS model is in agreement with GR  to describe TLNs for Schwarzschild black holes with masses $M \gtrsim 4.3 \times 10^4 M_{\textrm{Pl}}$. The constraint is stronger than existing ones in the literature and shows that the AOS model has a wider range of validity than expected so far. These findings reveal that, while classical black holes remain rigid in response to an external tidal field, the loop-quantized black holes exhibit quantum deformability, attributed to the inherent fuzziness of spacetime geometry. This indicates the presence of an additional macroscopic parameter of quantum origin for black holes, which serves as evidence for the existence of a quantum hair, a distinctive feature solely due to quantum gravitational effects. From a phenomenological perspective, the negative sign of TLNs, which has also been found for extended gravitational theories \cite{Cardoso:2017cfl, Cardoso:2018ptl, DeLuca:2022tkm, Barura:2024uog, Katagiri:2024fpn}, can also be a smoking gun to distinguish loop-quantized black holes from other black hole models.

The manuscript is structured as follows. In Section \ref{Section II}, we briefly review the AOS model and introduce the effective metric that captures the underlying loop quantum dynamics. In Section \ref{Section III}, we then define TLNs, outline the procedure for extracting TLNs by expanding the static response solutions at large distances, and present TLNs for the $D$-dimensional Schwarzschild black holes in GR, emphasizing that TLNs vanishes only in four dimensions in the classical theory. Section \ref{Section IV} is dedicated to calculating TLNs for the AOS model in response to an external tidal spin-$s$ field and demonstrating that they are all non-vanishing and negative for all multipoles. Finally, we conclude by discussing the implications of the non-vanishing of TLNs and providing remarks on our findings in \ref{Section V}. We use Planck units for discussion of results.

\section{Brief review of Ashtekar-Olmedo-Singh model} \label{Section II}

To quantize the interior region of a Schwarzschild black hole, the AOS model builds upon the well-established fact that the interior of a Schwarzschild black hole corresponds to a homogeneous cosmological model given by a Kantowski-Sachs geometry with symmetry group $\mathbb{R} \times SO(3)$. The homogeneity of the Kantowski-Sachs spacetime leads to a simple diagonal form for the Ashtekar-Barbero connection components ($b$, $c$) and conjugate triads ($p_{b}$, $p_{c}$), as a result of which, after fixing the Gauss constraint, the metric inside the Schwarzschild black hole in Kantowski-Sachs form reads as
\begin{align}\label{KS-metric-AB}
\mathrm{d} s^2 = - N^2 \mathrm{d} \tau^2 + \frac{p_b^2}{|p_c| L_o^2} \mathrm{d} x^2 + |p_c| (\mathrm{d} \theta^2 + \sin^2\theta \mathrm{d} \phi^2),
\end{align}
while $N$ is the lapse function and $p_{b}$ and $|p_{c}|$ are related to components of the standard Schwarzschild metric as
\begin{align}
    \frac{p_{b}^2}{|p_{c}| L_{o}^2} = \left(\frac{2m}{\tau}-1\right) , \ \ \ \ \ \ \ \ \ |p_{c}| = \tau^2,
\end{align}
where $m=G M$ with $M$ and $G$ being the mass of the black hole and the gravitational constant, respectively. $L_{o}$ is a fiducial length parameter introduced as an infrared cut-off to define the symplectic structure. $N,\; p_b$ and $p_c$ are also functions of $\tau$ only. The non-vanishing Poisson brackets among the phase space variables, $b$, $c$, $p_{b}$ and $p_{c}$, are given by
\begin{align}
    \{c, p_{c}\} = 2G\gamma,  \ \ \ \ \ \ \ \ \{b, p_{b}\} = G\gamma,
\end{align}
with $\gamma = 0.2375$ being the Barbero-Immirzi parameter fixed by black hole thermodynamics in LQG. Choosing $N = \gamma b^{-1} \textrm{sgn}(p_{c})p_{c}^{1/2}$, the classical Hamiltonian for the metric (\ref{KS-metric-AB}) can be written as follows
\begin{align}
\mathcal{H}_{\textrm{cl}}=-\frac{1}{2G\gamma} \left[2 c p_c +\left(b + \frac{\gamma^2}{b}\right) p_b \right].
\end{align}
In homogeneous and isotropic LQC, it is well established that a wide class of physical states are peaked at trajectories for the phase space variables that follow a dynamics generated by an effective Hamiltonian. It turns out that the effective Hamiltonian can be obtained from the classical Hamiltonian through the polymerization of connection variables $b$ and $c$ resulting in the following effective Hamiltonian constraint for the loop-quantized Kantowski-Sachs spacetime
\begin{align}
    H_{\mathrm{eff}} &=  - \frac{1}{2 G \gamma} \left[2 \frac{\sin (\delta_c c)}{\delta_c}  \, p_c  + \left(\frac{\sin (\delta_b b)}{\delta_b} + \frac{\gamma^2 \delta_b}{\sin(\delta_b b)} \right) \, p_b  \right], 
\end{align}
where $\delta_{b}$ and $\delta_{c}$ are regularization parameters of quantum origin. For the polymerized connection variables and for any chosen quantum parameters, by solving the effective dynamics, one can show that the black hole singularity is resolved, as expected from quantum gravity. However, the phenomenological predictions differ by how to choose the quantum parameters introduced to quantize the Hamiltonian constraint in detail. Different choices of $\delta_{b}$ and $\delta_{c}$ are frequently referred to as different schemes. There are three major schemes in the literature. The first one is the so-called $\mu_{0}$-like scheme where the quantum parameters are chosen to be a global constant in the phase space \cite{Ashtekar:2005qt, Modesto:2005zm}. The second one is the so-called $\bar{\mu}$-like scheme where the quantum parameters are chosen to be a function of phase space variables \cite{Chiou:2008nm}. Although both approaches resolve the interior singularity of the Schwarzschild black hole, they lead to several physically undesirable outcomes \cite{Ashtekar:2018cay}. To counter the problems of $\mu_{0}$-like and $\bar{\mu}$-like schemes, a new scheme was proposed in \cite{Ashtekar:2018lag} in which the quantum parameters are required to be Dirac observable. In fact, they are particular functions of the
phase space variables such that they are constant along the trajectories of the dynamical equations but may vary from one to another. Hence, the quantum parameters $\delta_{b}$ and $\delta_{c}$ in the large mass limit were found to be \cite{Ashtekar:2018lag, Ashtekar:2018cay}
\begin{align}\label{delta-bc}
    \delta_b = \left(\frac{\sqrt{\Delta}}{\sqrt{2\pi}\gamma^2 m}\right)^{\frac{1}{3}},  \ \ \ \ \ \ \ \
L_{o}\delta_c  = \frac{1}{2} \left(\frac{\gamma \Delta^2}{4\pi^2 m}\right)^{\frac{1}{3}}, 
\end{align}
where $\Delta = 4 \sqrt{3} \pi \gamma l_{\textrm{Pl}}^2$ is the minimum eigenvalue of the area operator in LQG and $l_{\textrm{Pl}}$ is the Planck length. For this choice of quantum parameters, it was found that the model exhibits the following desirable physical properties: i) The physical quantities and properties are independent of the infrared cut-off, i.e., the fiducial cell $L_{o}$; ii) The spacetime near the transition surface is symmetric; iii) There exists a universal mass-independent upper bound on the curvature invariants of the effective spacetime at the transition surface for large black holes; iv) No mass amplification of the white hole exists; v) The quantum corrections at the horizon scale are negligible. Therefore, the AOS model successfully replaces the interior singularity of the Schwarzschild black hole with a non-singular quantum bounce, addressing issues encountered in previous models.

In the AOS model one can extend the quantization to the exterior region of the Schwarzschild black hole, whose metric in Kantowski-Sachs form is given by
\begin{align}
\mathrm{d} s^2 =  - \frac{\tilde{p}_b^2}{|\tilde{p}_c| L_o^2} \mathrm{d} x^2 + N^2 \mathrm{d} \tau^2  + |\tilde{p}_c| (\mathrm{d} \theta^2 + \sin^2\theta \mathrm{d} \phi^2),
\end{align}
where $\tilde{p}_{b}$ and $\tilde{p}_{c}$ are conjugate triads for the exterior region. One can again find the exact solution for phase space variables from Hamilton's equations. Given the solution, one can then express the metric in our familiar static, spherically symmetric form
\bq\label{spherical-symmetric-metric}
\mathrm{d}s^2 = - f(r)\mathrm{d}t^2 + \frac{1}{g(r)} \mathrm{d}r^2 + r^2 \mathrm{d}\Omega^2,
\eq
where the metric functions $f(r)$ and $g(r)$ for macroscopic black holes, i.e., $\epsilon \ll 1$, are well-approximated by
\bq \label{AOS metric}
f(r) = \left(\frac{r}{r_{H}}\right)^{2\epsilon} \left(1- \left(\frac{r_{H}}{r}\right)^{1+\epsilon}\right), \ \ \ \ \ \ \ \ \ \ \ g(r) = \left(1- \left(\frac{r_{H}}{r}\right)^{1+\epsilon}\right),
\eq
with $\epsilon \simeq \frac{1}{2} \gamma^2 \delta_{b}^2$ and $r_{H} = 2GM$ is the black hole horizon. From the metric functions (\ref{AOS metric}), one finds that as $r$ goes to infinity, $f(r)$ diverges. However, the AOS metric can still be an asymptotically flat Minkowski spacetime at large distances as was shown in Ref. \cite{Ashtekar:2020ckv}. Hence, we will use this form of the AOS metric to calculate TLNs in section \ref{Section IV}.

\section{ Tidal Love numbers for Classical Schwarzschild Black Hole \\ in D Dimensions} \label{Section III}

In this section, we briefly revisit the definition of TLNs and explain how to extract TLNs using linear perturbation theory for spherically symmetric spacetime and finally review TLNs for the Schwarzschild black hole in $D$-dimensions in GR, where $D$ denotes the spacetime dimension.  For the spherically symmetric metric (\ref{spherical-symmetric-metric}), one can show that the radial component of spin-$s$ field perturbations, namely scalar ($s=0$), vector ($s=1$), and axial gravitational field perturbations ($s=2$), is governed by a Schrödinger-like equation given by
\bq\label{Schrödinger}
\frac{\mathrm{d}^2\Psi(r)}{\mathrm{d}r_{\star}^2} + \left[\omega^2-V(r)\right] \Psi(r) = 0,
\eq
where the effective potential reads as \cite{Arbey:2021jif, Konoplya:2024lch}
\begin{equation}\label{effective potential}
V(r) = f
\left[
\frac{\ell (\ell + 1)}{r^2} 
+ \frac{1 - s}{r} \frac{(f(r)g(r))^{\prime}}{2f(r)}
+ \frac{s(s - 1)(g(r) - 1)}{r^2}
\right],
\end{equation}
where prime denotes the derivative with respect to $r$ while $s$ being the spin of the test field and $\ell\ge s$. Besides, $r_{\star}$ is the tortoise coordinate linked to the radial coordinate according
to
\bq \label{rstar}
\mathrm{d}r_{\star} = \frac{\mathrm{d}r}{\sqrt{f(r)g(r)}}.
\eq
It is important to note that, under the Regge-Wheeler gauge, axial gravitational field perturbations in four-dimensional GR result in the same effective potential as described in Eq. (\ref{effective potential}) with $s=2$ \cite{Katagiri:2023umb}. Hereafter, we use the effective potential (\ref{effective potential}) with $s=2$ to study the axial gravitational field perturbations.

Assuming that the amplitude of the field is small, and its wavelength is much larger than the horizon scale, the field is then described by a static linear perturbation theory of the Schwarzschild black hole. This means that TLNs can be obtained from the zero-frequency limit, i.e., $\omega=0$, of Eq. (\ref{Schrödinger}) with effective potential (\ref{effective potential}). Assuming asymptotic flatness, i.e., $f(r)\rightarrow 1$ and $g(1) \rightarrow 1$ for $r\rightarrow \infty$, one finds that $\Psi(r)$ can be expanded asymptotically as \cite{Katagiri:2023umb}
 \bq \label{TLNs}
 \Psi(r) \sim  \left(\frac{r}{r_{H}}\right)^{\ell+1} \left[1 + \mathcal{O}\left(\frac{r_{H}}{r}\right)\right] + \kappa_{\ell}^{s}  \left(\frac{r_{H}}{r}\right)^{\ell} \left[1 + \mathcal{O}\left(\frac{r_{H}}{r}\right)\right].
 \eq
The first term $(r/r_{H})^{\ell+1}$, which corresponds to the growing mode, denotes the external tidal field applied at spatial infinity, while the second term, which corresponds to the decaying mode, $(r_{H}/r)^{\ell}$
encodes the induced response of the black hole. At the linear order, a tidal deformation induced by external static tidal fields is quantified by $\kappa_{\ell}^{s}$, representing induced multipole moments. For instance, in the case of a binary system, TLNs can measure the magnitude of the tidal deformation of one body due to the
force exerted by the other. The coefficient $\kappa_{\ell}^{s}$ denotes TLNs for spin $s$ external tidal fields and its value depends on the specified background geometry, the spin of the external tidal field and multipole $\ell$. As in Newtonian gravity, TLNs are defined as the ratio of induced response over the external source. However, there is an intrinsic ambiguity associated with the Newtonian matching procedure due to the fact that there is no unique separation of growing/decaying modes. Obviously, since $\ell$ is integer, there is an overlap between the source series and the response contribution. One possible way to remove such an ambiguity is to analytically continue $\ell$ to a complex plane where decaying and growing modes do not overlap and TLNs can be defined uniquely \cite{LeTiec:2020spy}. TLNs defined in Eq. (\ref{TLNs}) obtained from asymptotic expansion of linear perturbations at large distances which are power-law in $r$. The coefficient is constant and independent of scale. However, in the presence of matter or backgrounds deviate from GR, it is not guaranteed that the asymptotic expansion of the radial component of perturbations takes simple power-law series. Instead, the asymptotic behaviors can include logarithmic terms at large distances \cite{Katagiri:2023umb}
 \bq \label{TLNs-log}
 \Psi(r) \sim  \left(\frac{r}{r_{H}}\right)^{\ell+1} \left[1 + \mathcal{O}\left(\frac{r_{H}}{r}\right)\right] + \kappa_{\ell}^{s}  \left(\frac{r_{H}}{r}\right)^{\ell} \left[\ln \left(\frac{r}{r_{H}}\right) + \mathcal{O}(1)\right]\left[1 + \mathcal{O}\left(\frac{r_{H}}{r}\right)\right].
 \eq
The prefactor of the linear response term $(r_{H}/r)^{\ell}$ includes a logarithmic term and indeed it is scale dependent. The coefficient of the logarithmic term is interpreted as classical renormalization group as discussed in Ref. \cite{Hui:2020xxx}. It has also been found that TLNs exhibit such logarithmic running for high multipole $\ell$ in extended theories of gravity \cite{Barura:2024uog, Katagiri:2023umb}. 

\begin{table} 
\begin{center}
\begin{tabular}{c c c c c c}
$\kappa_{\ell}^{s}$ \ \ \ \  & \ \ \ \ $D=4$ \ \ \ \ & \ \ \ \ $D=5$ \ \ \ \ & \ \ \ \ $D=6$ \ \ \ \ & \ \ \ \ $D=7$ \ \ \ \ & \ \ \ \ $D=8 $  \ \ \ \ \\ \hline
\hline
$\kappa^{0}_{0}$ \ \ \ \ & \ \ \ \ $0$  \ \ \ \ & \ \ \ \ $0$ \ \ \ \ & \ \ \ \ $0$ \ \ \ \  & \ \ \ \ $0$ \ \ \ \ & \ \ \ \ $0$  \ \ \ \ \\ \hline
$\kappa^{0}_{1}$ \ \ \ \ & \ \ \ \ $0$  \ \ \ \ & \ \ \ \ $0.0625$ \ \ \ \ & \ \ \ \ $0.1290$ \ \ \ \  & \ \ \ \ $0.0912$ \ \ \ \ & \ \ \ \ $0.0746$  \ \ \ \ \\ \hline
$\kappa^{0}_{2}$ \ \ \ \ & \ \ \ \ $0$ \ \ \ \  & \ \ \ \ $0$ \ \ \ \  & \ \ \ \ $-0.0553$ \ \ \ \ & \ \ \ \  $0.1250$ \ \ \ \  & \ \ \ \ $0.1944$  \ \ \ \   \\ \hline
$\kappa^{0}_{3} \ \ \ \ $ &  \ \ \ \  $0$  \ \ \ \  &  \ \ \ \ 
 $0.0044$  \ \ \ \  &  \ \ \ \  $0$  \ \ \ \   &  \ \ \ \ 
 $-0.0257$  \ \ \ \  &  \ \ \ \  $-0.1169$  \ \ \ \  
\end{tabular}
\caption{TLNs in response to the scalar field for a $D$-dimensional Schwarzschild black hole with different values of dimension $D$ and multipole $\ell$. We have factored out the logarithmic dependence for those cases where $L$ has a half integer value.}
\label{TLNs-d}   
\end{center}
\end{table}

From the metric (\ref{spherical-symmetric-metric}), it is obvious that the exponent of $(r_{H}/r)$ scales differently from GR. This is very similar to the Schwarzschild black hole in $D$-dimensions where $f(r) = g(r) = 1- ({r_{H}}/{r})^{D-3}$, although there is an extra factor of $(r/r_{H})^{2\epsilon}$ in $f(r)$ for the AOS model. Hence, it is useful to review TLNs for the $D$-dimensional Schwarzschild black hole before moving to the next section where we consider the AOS model.  Reverting from the tortoise coordinate $r_{\star}$ back to the original radial coordinate $r$, and in the zero frequency limit, i.e. $\omega = 0$, and dividing by $f(r)$, Eq. (\ref{Schrödinger}) for the Schwarzschild black hole in $D$-dimension and for scalar field perturbation reads as (see Ref. \cite{Hui:2020xxx} for details)
\begin{align}\label{Sch-perturbations}
    f(r) \Psi^{\prime\prime} (r)+ f^{\prime}(r) \Psi^{\prime}(r) - \left[\frac{\ell(\ell+D-3)}{r^2} + \frac{(D-4)(D-2)}{4r^2} + \frac{D-2}{2r}f^{\prime}(r)\right]\Psi(r) = 0.
\end{align}
It is convenient to introduce the dimensionless radial variable 
\bq\label{dimensionless-parameter}
u = \left(\frac{r_H}{r}\right)^{D-3},
\eq
such that the black hole horizon is now located at $u = 1$, while spatial infinity corresponds to $u = 0$. Performing a field redefinition
\bq\label{field-redefintion}
\Psi(r) = u^{\frac{2\ell + D-4}{2(D-3)}}R(u(r)),
\eq
Eq. (\ref{Sch-perturbations}) can be expressed as a standard hypergeometric equation as follows
\bq\label{hyp-eq}
u(1-u)\frac{d^2 R(u)}{du^2} + \left[c - (a+b+1)u\right] \frac{d R(u)}{du} - a b R(u) = 0 .
\eq
Here $a$, $b$ and $c$ are given by
\bq\label{hyp-param-d}
a = b =  L+1, \ \ \ \ \ c = 2(L+1)
\eq
where $L = \ell/(D-3)$. Notice that the parameters $a$, $b$ and $c$ satisfy the condition $a+b-c=0$. The solution to Eq. (\ref{hyp-eq}) for different values of $a$, $b$, and $c$ has been studied in Ref. \cite{Hui:2020xxx} and it was found that the scalar TLNs, i.e., $s=0$, are given by 
\begin{align}\label{ks-non-integer}
    \kappa^{0}_{\ell} =  \frac{2L+1}{2\pi} \frac{\Gamma(L+ 1)^4 }{\Gamma(2L+2)^2} \tan (\pi L),
\end{align}
if $L$ is neither integer nor half-integer, and it is given by
\begin{align}\label{ks-log}
    \kappa^{0}_{\ell} = -\frac{(-1)^{2L} (D-3)\Gamma(L+1)^{2}}{(2L)! (2 L+1)!\Gamma(-L)^2},
\end{align}
for the case that $L$ is half-integer. We should point out that $\kappa^{0}_{\ell}$ reported in Eq. (\ref{ks-log}) is the coefficient of the logarithmic term, and it is scale dependent \cite{Hui:2020xxx}. In fact, we have factored out the logarithmic dependence and taken the coefficient as the TLN consistent with the definition in Eq. (\ref{TLNs-log}). Moreover, an extra minus sign appears due to the definition of TLNs in Eq. (\ref{TLNs-log}) where the coefficient of $\ln \left(\frac{r}{r_{H}}\right)$ is TLNs. Finally, for $D=4$, $L=\ell$, and from Eq. (\ref{ks-non-integer}), one can find that scalar TLNs vanish for Schwarzschild black holes in four dimensions. Although we just discussed the scalar TLNs here, it has been demonstrated in Ref. \cite{Hui:2020xxx} that all scalar, vector, and tensor TLNs vanish for Schwarzschild black holes in four-dimensions. 

As an illustrative example, we recorded the scalar TLNs of $D$-dimensional Schwarzschild black holes for different dimension $D$ and multipole $\ell$ in Table \ref{TLNs-d} for the first four multipoles, $\ell=0$, $\ell=1$, $\ell=2$, and $\ell=3$. From the first row in Table \ref{TLNs-d}, one can see that for $\ell=0$, $L=0$, and the scalar TLNs vanish for any dimension. For higher multipoles, one observes that the scalar TLNs for $D$-dimensional Schwarzschild black holes can be either positive, negative, or zero, while their magnitude depends on both $D$ and $\ell$. Among these TLNs, the largest one is for $D=8$ and $\ell=2$, and the smallest one is for $D=8$ and $\ell=3$. Moreover, there are three cases, namely $D=5$ and $\ell=1$,  $D=7$ and $\ell=2$, as well as $D=5$ and $\ell=3$, which exhibit logarithmic dependence (running). The scalar TLNs also vanish for $D=5$ and $\ell=2$ as well as $D=6$ and $\ell=3$, since $L$ is an integer in these cases. These results demonstrate that black holes in higher dimensions exhibit different responses to an external tidal field manifesting itself in TLNs. In the next section, we will calculate TLNs for the AOS model and show that they are non-vanishing even in four dimensions entirely due to quantum gravitational effects.

\section{Tidal Love numbers for Ashtekar-Olmedo-Singh model}\label{Section IV}

As we discussed in the previous section, TLNs are vanishing for four-dimensional Schwarzschild black holes, while they have non-vanishing TLNs in higher dimensions. In fact, it has been widely recognized that the vanishing of TLNs for Schwarzschild black holes in four-dimensional spacetime is related to the so-called ladder symmetry \cite{symmetry of TLNs, Combaluzier-Szteinsznaider:2024sgb, Gounis:2024hcm}. This property is specific to spherically symmetric black holes in four dimensions, and any deviation from GR leads to a non-vanishing TLNs, which can potentially be used to distinguish extended theories of gravity or even quantum gravitational effects \cite{Cardoso:2017cfl, Brustein:2020tpg}. Hence, we expect that TLNs be non-vanishing for the AOS model, which encodes information on the internal structure of loop-quantized black holes measurable from the exterior region. To calculate TLNs for the AOS model, one needs to solve Eq. (\ref{Schrödinger}) with the effective potential given by Eq. (\ref{effective potential}), which in the static limit and $r$ coordinate, after dividing by $f(r)$ reads as
\bq\label{spin-s}
g(r) \Psi^{\prime\prime}(r) +  \frac{(f(r)g(r))^{\prime}}{2f(r)}\Psi^{\prime}(r) - \left[
\frac{\ell (\ell + 1)}{r^2} 
+ \frac{1 - s}{r} \frac{(f(r)g(r))^{\prime}}{2f(r)}
+ \frac{s(s - 1)(g(r) - 1)}{r^2}
\right] \Psi(r) = 0.
\eq
We should point out a subtlety regarding the Eq. (\ref{spin-s}) in case of axial gravitational perturbations for the effective metric (\ref{AOS metric}). To compute the axial tensor perturbations, one usually linearizes the Einstein equation. As we pointed out previously, in the classical GR, the axial gravitational field perturbations lead to the same effective potential given by Eq. (\ref{effective potential}) with $s=2$. On the other hand, in this case, the effective metric comes from the effective Hamiltonian constraint and it is not the solution of the Einstein equation with quantum corrections. However, it has been shown that quantum corrections can be effectively modeled as contributions from an anisotropic fluid’s energy-momentum tensor within GR. In the tetrad formalism, it has been proven in Ref. \cite{Chen:2019iuo} that the axial components of the perturbed energy-momentum tensor defined by an anisotropic fluid are zero. Hence, our analysis assumes that the axial gravitational perturbations for the effective metric (\ref{AOS metric}) are governed by the effective potential (\ref{effective potential}) with spin $s=2$ as in classical GR. Inserting $f(r)$ and $g(r)$ given by Eqs. (\ref{AOS metric}) into Eq. (\ref{spin-s}), we reach
\begin{align} \label{spin-s-AOS}
\nonumber \left(1-\left(\frac{r_{H}}{r}\right)^{1+\epsilon}\right) \Psi^{\prime\prime}(r) & + \frac{1}{r_{H}} \left(\frac{r_{H}}{r} \right) \left( \left(\frac{r_{H}}{r}  \right)^{1+\epsilon} + \epsilon\right)\Psi^{\prime}(r) \\ & - \frac{1}{r_{H}^2} \left(\frac{r_{H}}{r} \right)^2 \left[\ell (\ell + 1) + (1-s) \left((1+s)\left(\frac{r_{H}}{r}\right)^{1+\epsilon} + \epsilon\right)
\right] \Psi(r) = 0,
\end{align}
where we used the following relation
\bq
\frac{(f(r)g(r))^{\prime}}{2f(r)}= \frac{1}{r_{H}} \left(\frac{r_{H}}{r} \right) \left(\left(\frac{r_{H}}{r} \right)^{1+\epsilon} + \epsilon \right).
\eq
Introducing the dimensionless radial variable 
\bq\label{dimensionless-parameter}
v = \left(\frac{r_{H}}{r}\right)^{1+\epsilon},
\eq
and performing the following field redefinition
\bq\label{field-redefintion}
\Psi(r) = v^{\ell}W(v(r)),
\eq
one finds that $\Psi^{\prime}(r)$ and $\Psi^{\prime\prime}(r)$ transform as follows
\begin{align}
\Psi^{\prime}(r) & =  - \frac{(1+\epsilon)}{r_{H}}\left(\ell v^{\ell+\frac{1}{1+\epsilon}} W(v) + {v^{\ell+1 + \frac{1}{1+\epsilon}}} \frac{dW(v)}{dv}\right),\\
\nonumber \Psi^{\prime\prime}(r)  & = \frac{(1+\epsilon)^2}{r_{H}^2} \left[ \ell \left(\ell+ \frac{1}{1+\epsilon}\right) v^{\ell + \frac{2}{1+\epsilon}}W(v) + \left(2\ell+1+ \frac{1}{1+\epsilon}\right)v^{\ell+1+\frac{2}{1+\epsilon}} \frac{dW(v)}{dv} \right. \\ & \left. + v^{\ell+2+ \frac{2}{1+\epsilon}} \frac{d^2 W(v)}{dv^2}\right].
\end{align}
Inserting $\Psi^{\prime}(r)$ and $\Psi^{\prime\prime}(r)$ back into Eq. (\ref{spin-s-AOS}) and dividing by $ (1+\epsilon)^2 v^{\ell+1+ \frac{2}{1+\epsilon}}/r_{H}^2$, and collecting coefficients for $W(v)$, $dW(v)/dv$, and $d^2 W(v)/dv^2$, Eq. (\ref{spin-s-AOS}) takes the following form
\begin{align}
   \nonumber   v (1-
   v) \frac{d^2 W(v)}{dv^2} & + \left[2\left(\ell+\frac{1}{1+\epsilon}\right) - \left(2\left(\ell+\frac{1}{1+\epsilon}\right) + 1\right) v \right]\frac{dW(v)}{dv} \\ & -  \left[ \left(\ell + \frac{1+s}{1+\epsilon}\right) \left(\ell + \frac{1-s}{1+\epsilon}\right) - \frac{\epsilon  \left(\ell^2 (2 + \epsilon )-\ell \epsilon +s-1\right)}{(1+\epsilon )^2 v} \right] W(v)=0.
\end{align}
This equation is similar to standard Hypergeometric equation with an extra term in the coefficient of $W(v)$. To eliminate the extra term and transform the equation into standard Hypergeometric equation, we use $W(v) = v^{p} F(v)$. After collecting terms for $F(v)$, $dF(v)/dv$ and $d^2F(v)/dv^2$, we reach the following equation
\begin{align}
   \nonumber   v (1-v) \frac{d^2F(v)}{dv^2} & + \left[2\left(\ell + p +\frac{1}{1+\epsilon} \right) - \left(2\left(\ell + p +\frac{1}{1+\epsilon}\right) + 1\right) v \right]\frac{dF(v)}{dv} \\ & -  \left[ \left(\ell + p + \frac{1+s}{1+\epsilon}\right) \left(\ell + p + \frac{1-s}{1+\epsilon}\right) - \frac{A}{(1+\epsilon)^2 v} \right] F(v)=0,
\end{align}
where $A$ is given by
\begin{align}
    A= p^2 (1 + \epsilon )^2 + p (1+ \epsilon ) (2 \ell (1 + \epsilon )-\epsilon +1) + \epsilon  \left(\ell^2 (2 + \epsilon )-\ell \epsilon +s-1\right).
\end{align}
Requiring $A$ to vanish, we can get rid of the extra term in the coefficient of $F(v)$ for the following values of $p$
\begin{align}
    p = \frac{\pm \sqrt{4 \ell (\ell+1)+\epsilon  (\epsilon - 4s +2)+1}-2 \ell (1 + \epsilon )+\epsilon -1}{2 (1 + \epsilon )},
\end{align}
but in order to recover GR, one should choose the positive sign. Hence, the equation can be rewritten in the standard hypergeometric equation (\ref{hyp-eq}) for $F(v)$ with the following parameters
\bq\label{AOS-abc}
a = \tilde{L} + 1 + \frac{s}{1+\epsilon}, \ \ \ \ \ \ b = \tilde{L} + 1 - \frac{s}{1+\epsilon},  \ \ \ \ \ \ c = 2 \tilde{L} + 2,
\eq
with $\tilde{L}$ given by
\begin{align}
\tilde{L} & = \frac{\sqrt{4 \ell (\ell+1)+\epsilon  (\epsilon - 4s +2)+1}}{2 (1 + \epsilon )} - \frac{1}{2}.
\end{align}
The parameters given by (\ref{AOS-abc}) for $s=0$ are exactly the same as the parameters \eqref{hyp-param-d} for the classical Schwarzschild black hole with the new $\tilde{L}$, which depends on the quantum gravitational parameter $\epsilon$. For $\epsilon=0$, $\tilde{L}=\ell$, and we recover the equation governing perturbations for the classical Schwarzschild black hole in four dimensions. Since $\tilde{L}$ is neither an integer nor a half-integer due to the presence of the quantum parameter $\epsilon$, which is related to the polymerization parameter $\delta_{b}$ given by Eq. (\ref{delta-bc}), whereby $a$, $b$, and $c$ are neither an integer nor a half-integer. Therefore, there is no mixing between growing and decaying modes, and TLNs are defined uniquely without the need for analytical continuation. The solutions for hypergeoemtric equation with parameters, which are neither an integer nor a half-integer have been investigated in Ref. \cite{Hui:2020xxx}, and it was found that TLNs can uniquely be obtained as follows 
\begin{figure}
    \centering
    \includegraphics[width=0.47\linewidth]{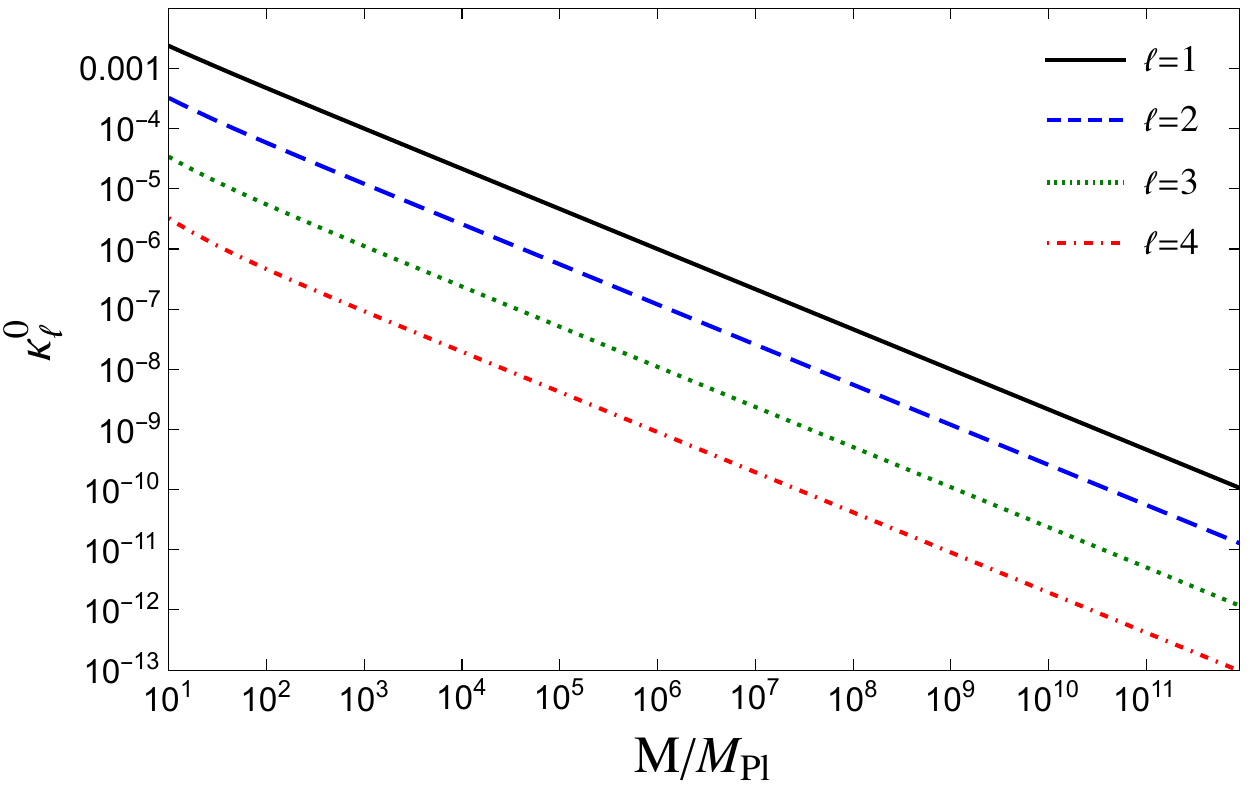}  \ \ \  \ \ \ \ \   \includegraphics[width=0.47\linewidth]{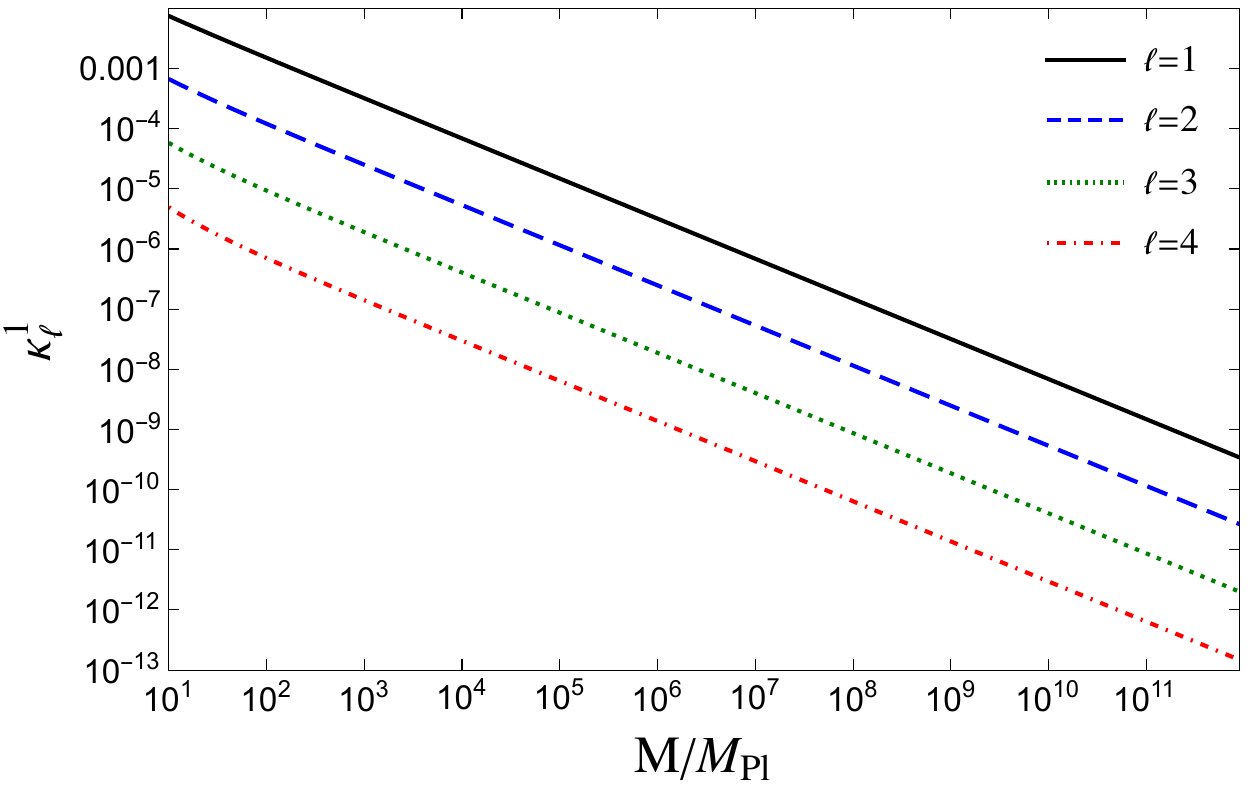}
    \caption{The magnitude of TLNs in response to a scalar field (left) and vector field (right) versus black hole mass $M$ for the first four multipoles $\ell$ in the case of the AOS black hole model.}
    \label{scalar-vector-love}
\end{figure}
\begin{align}
 \kappa_{\ell}^{s} = (a + b - 1)
\frac{ \Gamma(a)^2 \Gamma(b)^2}{\Gamma(c)^2} \frac{\sin(\pi a) \sin(\pi b)}{\pi \sin[\pi c]},
\end{align}
or equivalently, for the parameters $a$, $b$, and $c$ given by (\ref{AOS-abc}), reads as
\begin{align}\label{kls}
    \kappa_{\ell}^{s} =  (2\tilde{L}+1) \frac{\Gamma(\tilde{L}+ 1+ \frac{s}{1+\epsilon})^2 \Gamma(\tilde{L}+1- \frac{s}{1+\epsilon})^2}{\Gamma(2\tilde{L}+2)^2} \frac{\sin\left[\pi\left(\tilde{L} + \frac{s}{1+\epsilon}\right)\right] \sin\left[\pi\left(\tilde{L} - \frac{s}{1+\epsilon}\right)\right] }{\pi \sin (2 \pi \tilde{L})}.
\end{align}
For $\epsilon=0$, we have $\tilde{L} = \ell$ and $\kappa_{\ell}^{s}=0$, which is well-known result for the Schwarzschild black hole in four dimensions as we discussed in the previous section.

\begin{figure}
    \centering
    \includegraphics[width=0.47\linewidth]{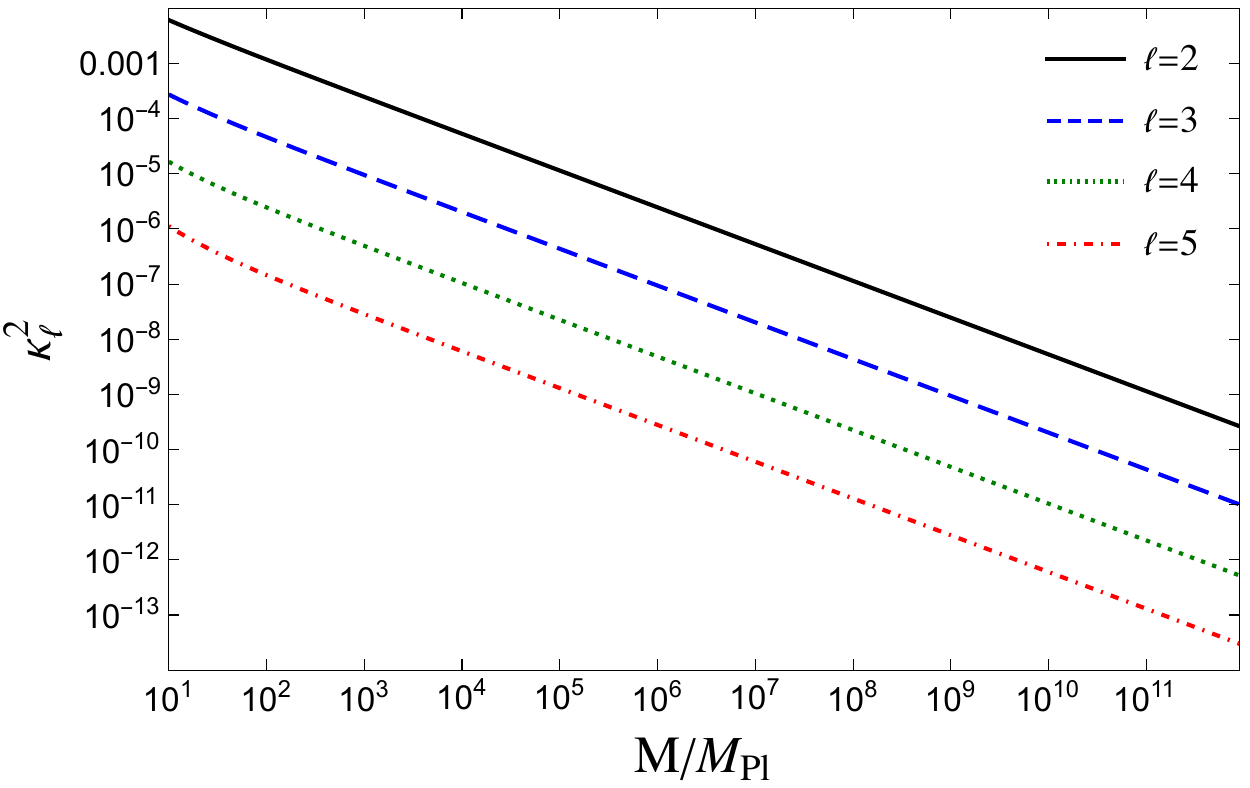}
    \caption{The magnitude of TLNs in response to an axial gravitational field versus $M$ for the first four multipoles $\ell$ in the case of the AOS black hole model.}
    \label{tensor-love}
\end{figure}

In Figs. \ref{scalar-vector-love} and \ref{tensor-love}, we plot the magnitude of TLNs ($|\kappa_{\ell}^{s}|$) for scalar, vector, and axial gravitational field responses versus black hole mass $M$ (in Planck units) for the first four multipoles in each case. The black solid, blue dashed, green dotted, and red dot-dashed curves represent the lowest to highest multipole values, respectively, with both axes plotted on a logarithmic scale. The black hole mass is considered in the range from  $10$ Planck masses to $10^{12}$ Planck masses. An interesting point is that all TLNs are non-vanishing and negative for all three scalar, vector, and axial gravitational field responses and all multipole $\ell$ (except for the scalar field response with $\ell=0$, for which $\kappa_{\ell = 0}^{s = 0} = 0$). Moreover, the magnitude of TLNs decreases as the mass of the black hole increases. Such Planck scale suppression is also reported in Ref. \cite{Kim:2020dif}, where quantum corrections to TLNs for Schwarzschild black holes are obtained from a sum rule for the semiclassical Hawking radiation spectrum. This means that TLNs are not detectable by the current generation of interferometers for astrophysical black holes. Moreover, we also find that for larger multipoles, TLNs have a smaller magnitude for the same black hole mass. Interestingly, we also found that TLNs for the AOS model do not exhibit logarithmic running for all three responses and any multipole $\ell$, which have appeared in the extended theories of gravity for large $\ell$ \cite{Barura:2024uog, Katagiri:2023umb}. 

The non-vanishing TLNs for loop-quantized AOS black holes suggest that, while classical black holes exhibit no deformability, the inherent quantum discreteness or fuzziness of spacetime geometry introduces quantum deformability in loop-quantized black holes. This indicates that black holes have a quantum hair related to their deformability in response to the external tidal field, which is accessible to outside observers. This also implies that the internal structure of loop-quantized black holes is fundamentally distinct from their classical counterparts. Regarding the negative sign of TLNs, we should point out that, using standard time-independent quantum perturbation theory, it has been shown that the quantum corrections to Love numbers are non-vanishing and generally negative \cite{Brustein:2020tpg}. Furthermore, the negative sign of TLNs has also been reported in modified gravity theories \cite{Katagiri:2024fpn} and also effective field theory of gravity \cite{Cardoso:2018ptl, Barura:2024uog}. This means that the non-vanishing TLNs can be used to phenomenologically distinguish loop-quantized black holes from other black hole models. Therefore, the non-vanishing TLNs due to quantum gravitational effects has profound theoretical and phenomenological implications.

\begin{figure}
    \centering
    \includegraphics[width=0.47\linewidth]{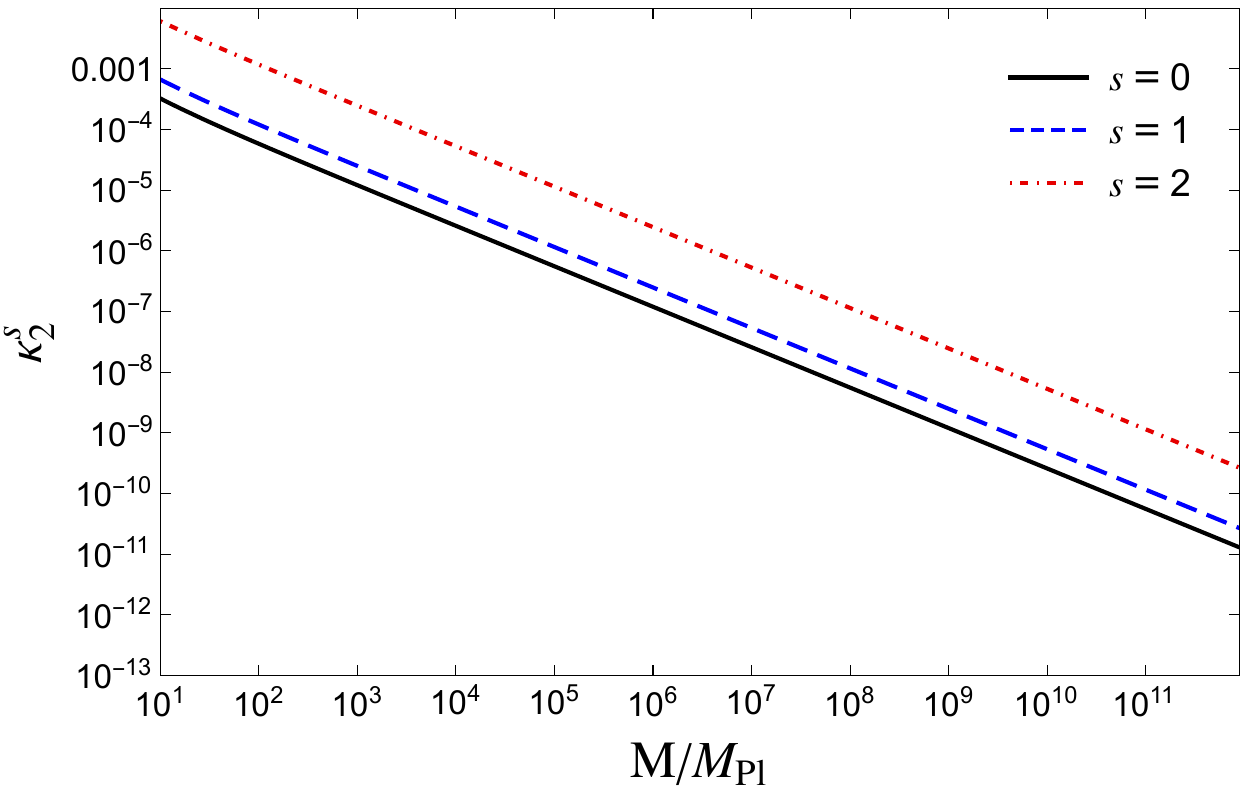}
    \caption{The magnitude of TLNs in response to scalar, vector, and axial gravitational fields versus black hole mass $M$ for quadrupole moment $\ell=2$ in the case of the AOS black hole model.}
    \label{scalar-vector-tensor-love}
\end{figure}

Additionally, we plot TLNs in response to the scalar (black solid curve), vector (dashed blue curve), and axial gravitational field (red dot-dashed curve) for the quadrupole moment, $\ell=2$, versus the mass of the black hole $M$ in Fig. \ref{scalar-vector-tensor-love}, from which one can observe that the magnitude of the tensor TLN is larger than the scalar and vector TLNs for the black hole mass. For instance, in the case of the largest black hole considered in Figs. \ref{scalar-vector-love} and \ref{tensor-love}, i.e., $M = 10^{12}M_{\text{Pl}}$, the magnitude of TLN in response to an axial gravitational field is approximately $\kappa_{s = 2}^{\ell=2} \simeq 10^{-10}$, while for scalar and vector responses, this is even one order of magnitude smaller. Finally, let us recall that polymerization parameters of the AOS model were derived in the large mass limit; hence, a larger correction to TLNs for Planckian mass black holes resulting in a larger deviation from GR is expected as it lies beyond the regime of validity of the AOS model. On the other hand, it has been found that the next generation of interferometers is potentially able to constrain quadrupole TLNs as small as $ \kappa_{\ell=2} \simeq 2\times 10^{-5}$. Hence, demanding that the AOS model be consistent with classical GR, one can place a lower bound on the black hole mass in the AOS model. In fact, in order to have non-detectable TLNs, the black hole mass in the AOS model should be $M \gtrsim 4.3 \times 10^{4} M_{Pl}$. In this sense, the constraint from vanishing TLNs for the AOS model points out that the AOS model has a wider range of validity than discussed so far in the literature.

\section{Conclusion}\label{Section V}

Tidal deformability of a self-gravitating object acted upon by an external long-wavelength
tidal field is characterized by a set of tidal Love numbers (TLNs) that encode information
about the object’s internal structure and dynamics of the gravitational field. Since the classical black hole spacetimes have singularities, and it is expected that these could be resolved by quantum gravity effects, a pertinent question is whether TLNs can encode information about residual quantum gravity effects in the exterior of black holes. Another way to pose this question is whether quantum gravitational black holes have the same rigid structure as their classical counterparts? Or do they deform when subjected to an external tidal field? 

To explore answers to these  questions, we investigated the static response of loop-quantized black holes, specifically the AOS model, to an external tidal spin-$s$ field, namely a scalar, a vector, and an axial gravitational field. We assumed the validity of the effective spacetime description as in various studies of loop quantum black holes and considered the static response of the effective metric to external tidal field perturbations. We found exact solutions for the equation governing the radial component of perturbations in terms of hypergeometric functions, which were then expanded at the large radial distances to extract TLNs. Since the parameters of the hypergeometric equation are all non-integer due to the presence of the quantum parameter, one of the polymerization parameters of the model, namely $\delta_{b}$, which is proportional to $M^{-1/3}$ where $M$ is the black hole mass, there is no mixing between growing and decaying modes, and TLNs are defined uniquely without the need for analytical continuation. Reading off TLNs from the ratio of the induced response over the external source, we found a closed form for TLNs as a function of the spin of the external tidal field $s$, multipole $\ell$, and quantum parameter of the model $\delta_{b}$. From Eq. (\ref{kls}), the main result of this manuscript, summarizing the leading-order loop quantum gravitational correction to Love numbers, is elegantly expressed as
\begin{align}\label{TLNs-leading-order}
    \kappa_{\ell}^{s} = - C (s, \ell) \left(\frac{M_{\textrm{Pl}}}{M}\right)^{\frac{2}{3}},
\end{align}
where $C(s, \ell)$ is a positive constant that depends on $s$ and $\ell$ with $\ell \ge s$. This aligns with physical intuition, as the Love number is a dimensionless quantity; the only dimensionless quantity that can be constructed from the only available free parameters of the model, the black hole mass $M$, and the fundamental scale of quantum gravity $M_{\textrm{Pl}}$, is the ratio $M_{\textrm{Pl}}/M$. However, the exact exponent and pre-factor should be fixed by the underlying theory of quantum gravity. 

From Eq. (\ref{TLNs-leading-order}), one immediately finds that all three scalar, vector, and tensor TLNs are non-vanishing and negative for all multipole $\ell$, and they are suppressed by the black hole mass (except for $s=0$ and $\ell=0$, which vanishes). We also observed that the magnitude of TLNs decreases for larger multipoles for all three responses with the same black hole mass. Our results also reveal that TLNs in response to the axial gravitational field have a larger magnitude in comparison with TLNs in response to scalar and vector fields. Additionally, we found that the non-detection of tidal deformability, which can potentially be measured up to $ \kappa_{\ell=2} = 2 \times 10^{-5}$, places a bound on the black hole mass in the AOS model below which there can be observable signatures. We showed that if one demands the mass of the black hole to be $M \gtrsim 4.3 \times 10^{4} M_{\textrm{Pl}}$, then no observable effect via TLNs emerges for the AOS model. Since the AOS model is primarily designed for the quantization of macroscopic black holes, our results show that it has a much wider range of validity.

Although the magnitude of TLNs for astrophysical black holes is not detectable by current technology, it has profound theoretical and phenomenological implications. From a phenomenological perspective, the negative sign of TLNs can be used to distinguish the AOS model from other loop-quantized black holes and also from other black hole solutions in the theories of modified gravity. Intuitively, the negative sign of TLNs indicates that the object deforms in a manner that counteracts the applied external tidal field, similar to how material polarizability works in materials. From a theoretical perspective, the non-vanishing loop quantum gravitational corrections to Love numbers manifest themselves as a form of quantum hair for loop-quantized black holes that is accessible to outside observers. This implies that loop-quantized black holes have a fundamentally distinct internal structure compared to classical black holes due to the inherent fuzziness of spacetime geometry. Hence, we conclude that although classical black holes are rigid in response to external tidal fields, loop-quantized black holes are quantum mechanically deformed.

Our analysis opens a new window to explore potential loop quantum gravitational effects in astrophysical phenomena. An immediate goal is to extend the study of TLNs for other loop-quantized black holes, especially those not necessarily based on the loop quantization of the Kruskal spacetime and which allow generalization beyond the homogeneous setting of the AOS model. 
Another aim is to distinguish the underlying quantum ambiguities arising in loop-quantized black holes via Love numbers and tidal deformability responses. Our preliminary investigations indicate that even for similar loop-quantized black holes, TLNs can leave distinct signatures \cite{second draft}. It would be useful to extend this analysis to the rotating version of loop-quantized black holes constructed from the non-rotating loop-quantized black holes \cite{Devi:2021ctm, Ban:2024qsa} through the Newman-Janis algorithm to compare the results with Kerr black holes to understand the effects of the spin of black holes on TLNs. It would be interesting to compute the dissipative TLNs (frequency-dependent response) and rigorously analyze the potential detection of the loop quantum gravitational effects through tidal effects imprinted in gravitational waves. With all these exciting avenues and more, we are hopeful that TLNs can serve as an important tool to restrict quantization ambiguities in loop-quantized models along the same lines as CMB has been used to restrict different loop quantizations of cosmological spacetimes (see for e.g. \cite{Li:2024xxz}).

\section*{Acknowledgments}
 
This work is supported by NSF grants PHY-2110207 and PHY-2409543.

\end{document}